\documentstyle[12pt]{article}
\begin{document}

\begin{center}
\vskip 2.5cm

{\Large \bf
{Gravitational Trapping for Extended Extra Dimension}
}\\
\vskip 2.5cm
{ Merab GOGBERASHVILI } \\
\vskip 2.5cm
{\it
 {Institute of Physics, Georgian Academy of Sciences}\\
 {6 Tamarashvili Str., Tbilisi 380077, Georgia}\\
 {(E-mail: gogber@hotmail.com)}} \\
\vskip 2.5cm

{\Large \bf Abstract}\\
\vskip 2.5cm
\end{center}

The solution of Einsteins equations for 4-brane embedded in
5-dimensional Anti-de-Sitter
 space-time is found. It is shown that the cosmological constant can
provide the existence
of ordinary 4-dimensional Newton's low and trapping of a matter on the
brane.

\vskip 2.5cm


\newpage

In conventional Kaluza-Klein's picture extra dimensions are curled up to
unobservable size. Recently multidimensional models with macroscopic
extra dimensions become popular \cite{ADD}. However, till the present
time little attention is paid to multidimensional models with extended
extra dimensions, where Universe is considered as a thin membrane in a
large-dimensional hyper-Universe \cite{RS,V,S,BK,G1,G2,G3}. This
approach also do not contradict to present time experiments \cite{OW}.

In the articles \cite{G1,G2} we had considered the model of Universe as
a 3-bubble expanding in 5-dimensional space-time. Two observed facts of
modern cosmology, the isotropic runaway of galaxies and the existence of
a preferred frame in Universe where the relict background radiation is
isotropic, have the obvious explanation in those models.

Here for the simplicity we again restrict ourselves with the case of
five dimensions. The general procedure immediately generalizes to
arbitrary dimensionality.

In models with non-compact extra dimensions one needs a mechanism to
confine a matter and to explain observed Newton's conventional law
inside of 4-dimensional manifold. It is natural to explain this trapping
as a result of the special solution of 5-dimensional Einstein's
equations \cite{V,G1,G2,G3}. The trapping has to be gravitationally
repulsive in nature and can be produced, for example, by large
5-dimensional cosmological constant \cite{V,G1}.

As it was shown in our paper \cite{G3} the metric tensor, corresponding
to the stable splitting of 5-dimensional space-time, slightly
generalizes the standard Kaluza-Klein one
\begin{equation}  \label{1.1}
g_{\alpha \beta }=\lambda ^2(x^5)\eta_{\alpha \beta
}(x^\nu)~~,~~~g_{55}=-1~~,~~~g_{5\beta }=0~~.
\end{equation}
Here $x^\nu$ are ordinary coordinates of 4-dimensional space-time,
$\eta_{\alpha \beta }(x^\nu)$ is 4-dimensional metric tensor and
$\lambda ^2(x^5)$ is the arbitrary function of fifth coordinate.
Solution (\ref{1.1}), which in \cite{G3} was received from the stability
conditions, exactly coincides with the anzats of Rubakov-Shaposhnikov
\cite{RS}.

In this paper we find exact solution of Einstein's equations for
4-brane. This solution is responsible for gravitational trapping of a
matter and for the hiding of 4-dimensional cosmological constant
resulting effective 4-dimensional Newton's law on the brane.

Simple demonstration of the gravitational trapping and Newton's law
restoring mechanism was made by us in the paper \cite{G1}. Because of
importance of this question we want to repeat here some results of this
paper.

As it will be shown below zero-zero component (the only component we
need now) of the metric tensor (\ref{1.1}) has the form
\begin{equation}
{g}_{00} = \eta_{00}(x^\mu) e^{E^2|x^5|}~~.
\label{2.1}
\end{equation}
Factor $exp(E^2 |x^5|)$ here, which rapidly increases fare from 4-brane
$x^5 = 0$, is responsible for the gravitational trapping of a matter.

The integration constant $E^2$ (corresponding to the width of
4-dimensional world $\epsilon$) in (\ref{2.1}) must be taken
proportional to 5-dimensional cosmological constant
\begin{equation}
E^2 \sim \Lambda^{1/2} \sim 1/\epsilon ~~,
\label{2.2}
\end{equation}
to live 4-dimensional world without the cosmological constant.

Einstein's 5-dimensional equations with the cosmological term for the
trapped point-like source in Newton's approximation gives
\begin{equation}
(\Delta - \Lambda ){g}_{00} = 6\pi^2 G M \delta (r) \delta (x^5) ~~,
\label{2.3}
\end{equation}
where $\Delta$ is the 4-Laplacian (including derivatives with respect to
$x^5$). Using (\ref{2.1}) and (\ref{2.2}) and separating variables we
obtain ordinary 4-dimensional Newton's formula on the brane without the
cosmological term
\begin{equation}
\eta_{00} = 1 - 2gM/r ~~,
\label{2.4}
\end{equation}
where
\begin{equation}
g \sim G/\epsilon
\label{2.5}
\end{equation}
is 4-dimensional gravitational constant \cite{G1}.

At the distances of the brane thickness in the right hand side of
equation (\ref{2.3}) we must put 4-dimensional delta function
$\delta(R)$, where $R$ is the radial coordinate in 5-dimensional
space-time. So we can't separate the variables and also hide the
cosmological constant $\Lambda$. The solution of (\ref{2.3}) in this
case is
\begin{equation}
{g}_{00} = \frac{2}{z} \left[I_1(z) - \Lambda GM~K_1(z)\right]~~,
\label{2.6}
\end{equation}
where $z= \Lambda^{1/2}R $  and $I_1, K_1$ are modified Bessel functions
of the order one. In the limit $z \ll 1$ (it means at the distances of
the branes width)
\begin{equation}
{g}_{00} = 1 - 2GM/R^2
\label{2.7}
\end{equation}
and thus 5-dimensional Newton's law is restored \cite{G1}.

Now we want to consider full system of nonlinear Einstein's equations
for the brane with the energy-momentum tensor
\begin{equation}
T_{\mu\nu} = - g_{\mu\nu}\sigma \delta (x^5), ~~~T_{55} = T_{\mu5} =
0~~,
\label{3.1}
\end{equation}
embedded in 5-dimensional Anti-de-Sitter space-time. Here $\sigma$ is
the tension of the brane.

As it was mentioned above (\ref{1.1}) 5-dimensional metric of Universe
in Gausian normal coordinates could be written in the form
\begin{equation}  \label{3.2}
ds^2=-(dx^5)^2+\lambda ^2(x^5) \eta_{\alpha \beta }dx^\alpha dx^\beta
~~.
\end{equation}
In these coordinates components of Christoffel's symbol with two or
three indices $5$ are equal to zero, while with the one index $5$ forms
the extrinsic curvature tensor \cite{MWT}
\begin{eqnarray}
\nonumber K_{\alpha\beta} = \Gamma^{5}_{\alpha\beta } =
\frac{1}{2}\partial_{5}g_{\alpha\beta } =
\lambda\lambda^{^{\prime}}\eta_{\alpha\beta }~~, \\
\label{3.3}K^{\alpha\beta} = - \frac{1}{2}\partial_{5}g^{\alpha\beta }
~~.
\end{eqnarray}
Prime denotes derivative with the respect to the coordinate $x^5$.

Also we would like to represent here some useful relations
\begin{eqnarray}  \label{3.4}
g^{\alpha\gamma}K_{\gamma\beta} = \Gamma^{\alpha}_{5\beta } =
\lambda\lambda^{^{\prime}}\delta^{\alpha}_{\beta }~~,  \nonumber \\
K = g^{\alpha\beta}K_{\alpha\beta} = g_{\alpha\beta}K^{\alpha\beta} =
4\lambda^{^{\prime}}/\lambda~~, \\
\partial_5 K= g^{\alpha\beta}\partial_5 K_{\alpha\beta} -
2K^{\alpha\beta}K_{\alpha\beta}~~.  \nonumber
\end{eqnarray}

Any vector or tensor is naturally split-up into its components
orthogonal and tangential to the brane. Using decomposition of the
curvature tensor
\begin{eqnarray}  \label{3.5}
^5R_{\alpha\beta} = R_{\alpha\beta } + \partial_5 K_{\alpha\beta } -
2K_{\alpha}^{\gamma} K_{\gamma\beta } + K K_{\alpha\beta }~~,  \nonumber
\\
^5R_{55} = - \partial_5 K - K^{\alpha\beta}K_{\alpha\beta}~~, \\
^5R = R + K^{\alpha\beta}K_{\alpha\beta} + K^2 + 2\partial_5 K
\nonumber
\end{eqnarray}
one can find decomposition of Einstein's equations
\begin{eqnarray}  \label{3.6}
R_{\alpha\beta} - \frac{1}{2}\eta_{\alpha\beta}R -
3\eta_{\alpha\beta}(\lambda \lambda ^{"} + \lambda ^{^{\prime }2}) = -
\eta_{\alpha\beta}\lambda ^2 [\Lambda+ 6\pi ^2G \sigma \delta (x^5)]~~,
\nonumber \\
R + 12\lambda ^{\prime ^2} = 2\lambda ^2 \Lambda ~~ .
\end{eqnarray}

In four dimensions to have Einstein's equations without the cosmological
term
\begin{eqnarray}  \label{3.7}
R_{\alpha\beta} - \frac{1}{2}\eta_{\alpha\beta}R = 0 ~~, \nonumber \\
R = 0 ~~,
\end{eqnarray}
we must put
\begin{eqnarray}  \label{3.8}
 3(\lambda \lambda ^{"} + \lambda ^{\prime ^2}) = \lambda ^2[\Lambda +
6\pi ^2G\sigma \delta (x^5)]~~, \nonumber \\
6\lambda ^{\prime ^2} = \lambda ^2 \Lambda ~~.
\end{eqnarray}
Using formula
\begin{equation}  \label{3.9}
|x^5|^{\prime} = H(x^5) - H(-x^5)~~,
\end{equation}
where $H(x^5)$ is the step function, one can show that system
(\ref{3.8}) has the trapping solution
\begin{equation}  \label{3.10}
\lambda = e^{E^2|x^5|}~~,
\end{equation}
where the integration constant has the value
\begin{equation}  \label{3.11}
E^2=\sqrt{\Lambda/6} = \pi^2G\sigma ~~.
\end{equation}
This formula also contains necessary relation between the brane tension
$\sigma $ and 5-dimensional cosmological constant $\Lambda $.

So we received the solution we have used in (\ref{2.1}) to demonstrate
our mechanism of gravitational trapping and Newton's law restoration on
the brane.

\end{document}